\documentclass[twocolumn,notitlepage,prl,superscriptaddress,longbibliography]{revtex4-2}
\usepackage{amsfonts}
\usepackage{textcomp}
\usepackage{times}
\usepackage{graphicx}
\usepackage{float}
\usepackage{latexsym,amsmath,amssymb,bm,euscript}
\usepackage{color}
\usepackage{subfigure}
\usepackage{epstopdf}
\usepackage[colorlinks=true,linkcolor=blue,citecolor=blue]{hyperref}
\usepackage{hyperref}
\usepackage{soul}
\usepackage[normalem]{ulem}
\usepackage{mathrsfs}
\usepackage{amsmath}
\usepackage{xspace}
\usepackage{natbib}
\usepackage{ulem}
\usepackage{etoolbox}
\usepackage{tikz}
\usepackage{physics}
\usepackage{multirow}
\usepackage{pifont}
\usepackage{tabularx}
\newcolumntype{Y}{>{\centering\arraybackslash}X}

\usepgflibrary {shadings}
\usetikzlibrary {arrows.meta}

\renewcommand{\Tr}[1]{\mathrm{Tr}\left\{#1\right\}}
\def\thetitle{Fluctuating Pair Density Wave in Finite-temperature Phase Diagram of the $t$-$t^\prime$ Hubbard Model}

\definecolor{DarkBlue}{RGB}{50,120,180}
\definecolor{TextBlue}{RGB}{30,70,150}

\begin{document}
\title{\thetitle}

\author{Qiaoyi Li}
\affiliation{Institute of Theoretical Physics, Chinese Academy of Sciences, Beijing 100190, China}
\affiliation{School of Physical Sciences, University of Chinese Academy of Sciences, Beijing 100049, China}

\author{Yang Qi}
\affiliation{State Key Laboratory of Surface Physics and Department of Physics, Fudan University, Shanghai 200433, China}
\affiliation{Hefei National Laboratory, Hefei 230088, China}

\author{Wei Li}
\email{w.li@itp.ac.cn}
\affiliation{Institute of Theoretical Physics, Chinese Academy of Sciences, Beijing 100190, China}
\affiliation{Hefei National Laboratory, Hefei 230088, China}

\begin{abstract} 
The Hubbard model and its extensions are canonical theoretical frameworks for understanding correlated electronic states, including those in high-$T_c$ cuprates. Here, we use state-of-the-art thermal tensor network method to map out the temperature-doping phase diagram of the $t$-$t'$ Hubbard model. On the electron-doped side, we find a $d$-wave superconducting (dSC) regime, supporting the scenario of high-$T_c$ superconductivity. In contrast, on the hole-doped side, no robust dSC phase is detected. Instead, a finite-temperature regime dominated by strong pair-density-wave (PDW) fluctuations emerges, which may eventually give way to charge density wave order upon further cooling. The PDW state exhibits inter-arc pairing with net momentum near $(0, \pi)$, distinct from the zero-momentum pairing in conventional dSC. Furthermore, these fluctuating PDW states occupy the lower portion of the pseudogap regime on the hole-doped side. We provide a comprehensive finite-temperature perspective consistent with previous ground-state studies, shedding new light on pairing instabilities and exotic electronic states in high-$T_c$ superconductors. 
\end{abstract}
\date{\today}

\maketitle

\textcolor{blue}{\textit{Introduction}}.--- The high-$T_{\rm c}$ $d$-wave superconductivity (dSC) in cuprates~\cite{Dagotto1994RMPCorrelated, Lee2006RMPDoping, Keimer2015Nquantum} remains a long-standing and intriguing question in condensed matter physics. The Hubbard model~\cite{Hubbard1963PRSAElectron} and its variant, the $t$-$J$ model~\cite{Zhang1988PRBEffective}, have served as important theoretical platforms for understanding correlated electrons, particularly dSC, in cuprate superconductors and beyond~\cite{Qin2022ARoCMPHubbard}. By including both nearest-neighbor (NN) hopping $t$ and the next-nearest-neighbor (NNN) hopping $t^\prime$, single-band $t$-$t'$ Hubbard or $t$-$t'$-$J$ model has been shown to capture certain key features of the cuprate phase diagram: the antiferromagnetic (AFM) Mott insulator at half filling~\cite{Schaefer2021PRXTracking, Wietek2021PRXMott, Qu2024PRLPhase}, existence of dSC dome~\cite{Maier2005PRLSystematic, Sordi2012PRLStrong, Gull2013PRLSuperconductivity, Jiang2018PRBSuperconductivity, Jiang2019SSuperconductivity, Chung2020PRBPlaquette, Jiang2020PRRGround, Jiang2024PRBGround, Xu2024SCoexistence, Zhang2025PRLFrustration, Jiang2018PRBSuperconductivity, Jiang2021PotNAoSGround, Gong2021PRLRobust, Lu2024PRLEmergent, Qu2024PRLPhase, Chen2025PNASGlobal} and charge density waves (CDW)~\cite{Zheng2017SStripe, Huang2018nQMStripe, Ponsioen2019PRBPeriod, Jiang2020PRRGround, Qin2020PRXAbsence, Wietek2021PRXStripes, Mai2022PotNAoSIntertwined, Xu2022PRRStripes, Mai2023NCRobust, Xiao2023PRXTemperature, Jiang2024PRBGround, Shen2024ground, Xu2024SCoexistence, Liu2025PRLAccurate} at finite doping, along with pseudogap (PG)~\cite{Senechal2004PRLHot, Macridin2006PRLPseudogap, Sordi2012PRLStrong, Gull2013PRLSuperconductivity, Gunnarsson2015PRLFluctuation, Gull2015PRBQuasiparticle, Chen2017NCSimulation, Wu2018PRXPseudogap, Schaefer2021PRXTracking, Wietek2021PRXStripes, Wu2022PotNAoSNon, Simkovic2024SOrigin, Stepanov2025Superconductivity} and the strange-metal (SM)~\cite{Huang2019SStrange, Wu2022PotNAoSNon} states.

Despite significant advances in numerical simulations, there are open questions remaining to be resolved. A prime example is the superconductivity in the doped $t$-$t'$ Hubbard model: while constrained-path auxiliary-field quantum Monte Carlo (CP-AFQMC) studies~\cite{Xu2024SCoexistence} suggest a stronger dSC order on the hole-doped side than that in the electron-doped side, density matrix renormalization group (DMRG)~\cite{Jiang2024PRBGround, Jiang2025Competition} and infinite projected entangled-pair state (iPEPS)~\cite{Zhang2025PRLFrustration} calculations report that the hole-doped dSC is weaker or even absent. The weaker dSC on the hole-doped side is also observed in the $t$-$t^\prime$-$J$ model~\cite{Jiang2021PotNAoSGround,Jiang2022PRBPairing, Wietek2022PRLFragmented, Lu2024PRLEmergent, Qu2024PRLPhase}, in sharp contrast to most experiments on cuprates~\cite{Lee2006RMPDoping, Sobota2021RMPAngle}. Whether this inconsistency originates from differences in numerical techniques --- such as finite-size effects or the constrained-path approximation~\cite{Zhang1999PRLFinite, He2019PRBFinite} --- or points to an intrinsic limitation of single-band models remains a key question to be addressed.

While most studies of the $t$-$t'$ Hubbard model have focused on ground‑state properties, its finite‑temperature pairing behavior remains largely unexplored. Determinant quantum Monte Carlo (DQMC) studies have provided insights in certain regimes~\cite{Blankenbecler1981PRDMonte, Scalapino1981PRBMonte, Hirsch1983, Assaad2008, Wang2025Finite}, but they remain constrained to limited regions of the temperature–doping phase diagram.
In this work, we perform large-scale thermal tensor-network (ThermoTN) simulations of the $t$-$t^\prime$ Hubbard model with the tangent-space tensor renormalization group (tanTRG) method~\cite{Li2023PRLTangent,Li2026PRBThermal}. 
This approach allows us to obtain the phase diagram covering both electron- and hole-doping regimes up to $|\delta| = 0.2$ and down to ultralow temperature of $T/t \simeq 0.02$. At low temperatures, our results are consistent with previous DMRG studies, confirming a dSC phase in the electron-doped regime and a CDW phase in the hole-doped regime \cite{Jiang2024PRBGround, Jiang2025Competition}. 

Remarkably, above the CDW ground states at hole doping, we uncover fluctuating pair density wave (PDW$^*$)~\cite{Lee2014PRXAmperean, Fradkin2015RMPColloquium, Agterberg2020ARoCMPPhysics} states at finite temperature, where the finite-momentum pairing with $\mathbf{Q}_{\rm PDW} \approx (0, \pi)$ exhibits a strength substantially exceeding that of conventional $d$-wave pairing.
We attribute the distinct pairing instabilities to the node-antinode structure of Fermi surface (FS). At finite temperature, the electron-doped side shows a suppression of single-particle spectral weight near the nodal points. In contrast, the hole-doped spectra display clear Fermi arcs with spectral weight concentrated near the nodes. Here, PDW fluctuations feature inter-arc pairing, which competes with the anticipated zero-momentum $d$-wave pairing. These results on the $t$-$t^\prime$ Hubbard model shed new light on pairing instabilities in cuprate superconductors.

\textcolor{blue}{\textit{Model and methods.}}---
We study the Hubbard model on a square lattice, defined by the Hamiltonian
\begin{equation}
     H = -\sum_{ij\sigma}t_{ij}c_{i\sigma}^\dagger c_{j\sigma} + U\sum_i n_{i\uparrow}n_{i\downarrow} - \mu\sum_i n_i,
\end{equation}
where $c_{i\sigma}^\dagger$ ($c_{i\sigma}$) creates (annihilates) an electron at site $i$ with spin $\sigma = \uparrow, \downarrow$. 
Here $n_{i\sigma} = c_{i\sigma}^\dagger c_{i\sigma}$ denotes the spin-resolved density, $n_i = n_{i\uparrow} + n_{i\downarrow}$ is the total electron density, and $t_{ij}$ includes NN ($t$) and NNN ($t^\prime$) hopping amplitudes. We set $t=1$ as the energy unit, and adopt fixed $t^\prime = -0.2$ and $U = 8$ that are relevant to cuprates~\cite{Hirayama2018PRBAb, Hirayama2019PRBEffective}.

To simulate the finite-temperature properties of the $t$-$t'$ Hubbard model, we employ the thermoTN approach~\cite{Wang1997PRBTransfer, Verstraete2004PRLMatrix, Feiguin2005PRBFinite, White2009PRLMinimally, Stoudenmire2010NJoPMinimally, Wietek2021PRXStripes, Czarnik2012PRBProjected, Czarnik2014PRBFermionic, Czarnik2016PRBVariational, Sinha2022PRBFinite, Li2011PRLLinearized, Dong2017PRBBilayer, Chen2017PRBSeries, Chen2018PRXExponential, Chen2021PRBQuantum}, in particular the tanTRG method~\cite{Li2023PRLTangent, Li2026PRBThermal}. By representing the thermal density operator $\rho(\beta) \equiv e^{-\beta H}$ as a matrix product operator (MPO)~\cite{Verstraete2004PRLMatrix, Dong2017PRBBilayer}, we prepare the MPO via imaginary-time evolution using time-dependent variational principle (TDVP)~\cite{Haegeman2011PRLTime, Haegeman2016PRBUnifying}, providing accurate simulations on cylinders~\cite{Li2023PRLTangent, Wang2023PRLPlaquette, Qu2024PRLBilayer, Qu2024PRLPhase}.

In practice, we employ cylindrical geometries with open boundaries along the length ($x$) direction, and combine periodic (PBC) and antiperiodic (APBC) boundary conditions along the width ($y$) direction to improve momentum resolution and mitigate finite-width effects~\cite{Lin2001PRETwist, Xu2024SCoexistence}.
Accelerated by the U(1)$_{\rm charge}\times$SU(2)$_{\rm spin}$ symmetry~\cite{Devos2025TensorKit.jl}, controlled bond expansion technique~\cite{Gleis2023PRLControlled, Li2024PRLTime} and interaction-based parallelism~\cite{Li2026FiniteMPS.jl}, we retain MPO bond dimension up to $D = 32768$ (see Fig.~\ref{Fig:EM_extrap} and Supplemental Material~\cite{Supplementary} for data convergence), enabling accurate simulations deeply into the low-temperature regime where the DQMC method suffers a sign problem~\cite{Iglovikov2015PRBGeometry}. Specifically, our simulations reach a space-time volume of $W\times L \times \beta=8\times18\times32 \simeq 4600$, substantially exceeding the largest size of about $8\times 8 \times 5 = 320$ in DQMC calculations~\cite{Wang2025Probing, Wang2025Finite}.

\begin{figure}[tbp]
\includegraphics[width = \linewidth]{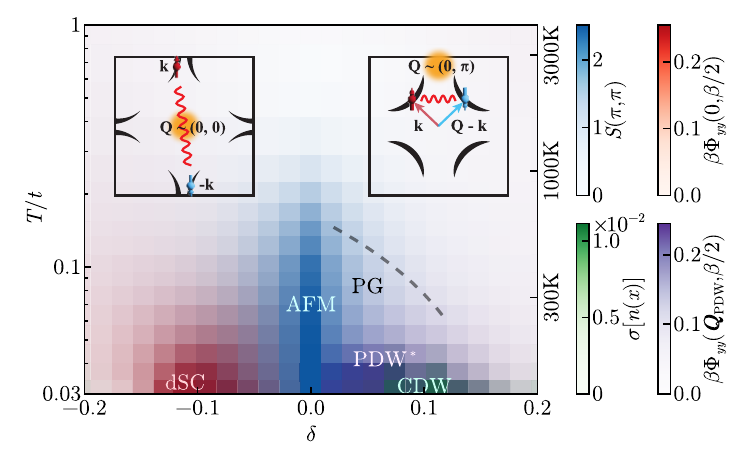}
\caption{Temperature-doping phase diagram of the $t$-$t^\prime$ Hubbard model with $t^\prime = -0.2$ and $U = 8$ on a $6\times18$ cylinder, obtained by averaging over PBC and APBC. A logarithmic temperature scale is used to highlight the low‑temperature behavior. The right axis indicates the corresponding temperature in Kelvin, obtained by assuming a typical cuprate hopping value of $t \simeq 4000$~K as energy unit. Typical pseudogap crossover temperatures are schematically indicated by the dashed line. The red and purple contours show $\beta \Phi_{yy}(\mathbf{q},\beta/2)$ at $\mathbf{q} = 0$ and $\mathbf{Q}_{\rm PDW} = (0, \pi)$, reflecting the $d$-wave SC and PDW fluctuations, respectively. The blue color scale represents the AFM correlation $S(\pi, \pi)$, while the green contour denotes $\sigma[n(x)]$, probing the charge inhomogeneity. The inset illustrates the node–antinode structure and the corresponding singlet‑pairing instabilities: (left) conventional zero‑momentum $d$‑wave pairing under electron doping, and (right) $(0,\pi)$‑PDW formed via inter‑arc pairing under hole doping.
}
\label{Fig1:PhaseDiagram}
\end{figure}

\textcolor{blue}{\textit{Imaginary-time proxy for spectral weight.}}---With the MPO representation of $e^{-\beta H / 2}$, the imaginary-time Green's function $G(\mathbf{k}, \beta/2) \equiv \expval{c_{\mathbf{k}\sigma}(\beta/2) \, c_{\mathbf{k}\sigma}^\dagger} = \Tr{e^{-\beta H/2}\, c_{\mathbf{k}\sigma} \, e^{-\beta H/2}\, c_{\mathbf{k}\sigma}^\dagger}/\Tr{e^{-\beta H}}$ can be efficiently computed. This quantity serves as an imaginary-time proxy (ITP) that encodes the single-particle spectral weight near the Fermi level ($\omega = 0$), as 
\begin{equation}
\beta G(\mathbf{k}, \beta/2) = \int_{-\infty}^\infty \frac{d\omega}{2\pi}\frac{\beta A(\mathbf{k}, \omega)}{2\cosh(\beta\omega/2)}.
\end{equation}
In the low-temperature limit, this kernel function $\beta / [2\pi\cosh (\beta\omega/2)]$ approaches the Dirac delta function, leading to an approximation $A(\mathbf{k}, \omega = 0) \approx 2\beta G(\mathbf{k}, \beta/2)$. Such approximation is widely used in DQMC studies~\cite{Lederer2017PNASSuperconductivity, Huang2019SStrange, Wang2021PRRNumerical, Jiang2022NCMonte, Wang2025Probing, Chen2025Topological} and has recently been integrated with the ThermoTN approach~\cite{Li2023PRLTangent, Qu2024PRLPhase, Xi2024Thermal, Gao2025Spin, Chen2026SBFractional, Qu2026$d$}. 

To characterize the momentum-resolved low-frequency pairing fluctuations, we also compute the imaginary-time correlator $\Phi_{\alpha\alpha}(\mathbf{q}, \beta/2) \equiv \expval{O_\mathbf{q}^\alpha(\beta/2) O_{-\mathbf{q}}^\alpha}$, with $O_\mathbf{q}^\alpha \equiv \frac{1}{\sqrt{N}}\sum_{i=1}^N e^{-i\mathbf{q}\cdot\mathbf{r}_i}O_{\alpha}(\mathbf{r}_i)$. The pairing operator is $O_\alpha(\mathbf{r}_i) = (\Delta_{\mathbf{r}_i, \mathbf{r}_i + \alpha} + {\rm h.c.})/2$, with $\alpha \in \{\hat{\mathbf{x}}, \hat{\mathbf{y}}\}$ and $\Delta_{ij} = (c_{i\downarrow}c_{j\uparrow} - c_{i\uparrow}c_{j\downarrow})/\sqrt{2}$.
The corresponding spectral representation reads
\begin{equation}
\beta \Phi_{\alpha\alpha}(\mathbf{q}, \beta/2) = \int_{-\infty}^\infty \frac{d\omega}{2\pi}\frac{\beta {\rm Im}\chi_{\alpha\alpha}(\mathbf{q}, \omega)}{\sinh(\beta\omega/2)},
\label{Eq:ITP_Phi}
\end{equation}
where $\chi_{\alpha\alpha}(\mathbf{q}, \omega)$ is the dynamical pairing susceptibility.
Similarly, taking low-temperature limit yields an approximation $\Phi_{\alpha\alpha}(\mathbf{q}, \omega = 0) \approx \frac{2\beta}{\pi}\Phi_{\alpha\alpha}(\mathbf{q}, \beta/2)$~\cite{Supplementary}, where $\Phi_{\alpha\alpha}(\mathbf{q}, \omega = 0)$ constitutes a zero-frequency pairing structure factor.
Moreover, the convolution kernel $\beta/\sinh(\beta\omega)/2 \sim e^{-\beta\omega/2}$ serves as a ``low-pass" filter, while the equal-time $\Phi_{\alpha\alpha}(\mathbf{q})$ involves high-frequency (and short-range) contributions [c.f. Eq.~\eqref{Eq:SR_all}].
Therefore, the former grants access to physically informative low‑frequency fluctuations that are expensive to obtain in conventional dynamical methods~\cite{Paeckel2019AoPTime, Kuehner1999PRBDynamical, Jeckelmann2002PRBDynamical, Holzner2011PRBChebyshev, Dargel2012PRBLanczosa, Kovalska2025Tangent}.

\textcolor{blue}{\textit{The temperature-doping phase diagram.}}--- In Fig.~\ref{Fig1:PhaseDiagram}(a), we present the temperature-doping phase diagram of the $t$-$t^\prime$ Hubbard model. The doping level $\delta = 1 - n$, where $n$ is the average electron filling, represents hole doping ($\delta > 0$) or electron doping ($\delta < 0$) relative to half-filling. To provide a comprehensive overview, we present a joint contour plot that simultaneously characterizes three key correlations: The pairing correlation $\beta \Phi_{yy}(\mathbf{q}, \beta/2)$, magnetic correlation $S(\mathbf{q}) = \frac{1}{3N}\sum_{ij}e^{-i\mathbf{q}\cdot(\mathbf{r}_i - \mathbf{r}_j)}\expval{\mathbf{S}_i\cdot \mathbf{S}_j}$, and standard deviation of local charge density $\sigma[n(x)]$~\cite{Jiang2025Competition}. To capture bulk behavior, the four outermost columns at each boundary of the cylinder are excluded from the calculation.

In Fig.~\ref{Fig1:PhaseDiagram}, we observe that strong AFM correlations near half‑filling weaken upon doping, albeit remaining stronger on the electron‑doped side compared to the hole‑doped side. Such particle–hole asymmetry is consistent with previous studies on the $t$-$t^\prime$-$J$ model~\cite{Qu2024PRLPhase} and the established phase diagram of cuprate superconductors~\cite{Dagotto1994RMPCorrelated, Lee2006RMPDoping}.
On the electron‑doped side, a dome-shaped dSC regime emerges near optimal doping $\delta_{\rm o} \approx -1/9$, once AFM correlations are suppressed. In contrast, on the hole‑doped side, pronounced CDW modulations appear, indicating a distinct competing order in that regime. These finite-temperature results are consistent with previous DMRG studies on the $t$-$t^\prime$ Hubbard model~\cite{Jiang2024PRBGround}, where a dSC phase is observed at electron doping and two charge ordered states, namely Wigner crystal and $2/3$-filled CDW, are found under hole doping on the width-6 cylinder.

Beyond ground-state studies, as indicated by the purple contour in Fig.~\ref{Fig1:PhaseDiagram}(a), a fluctuating PDW phase exists above the CDW phase at intermediate temperatures.
An exception occurs in the 4‑leg cylinder, where the $(0,\pi)$‑PDW can equivalently be viewed as a plaquette d-wave superconductivity, which retains till the ground state~\cite{Jiang2020PRRGround, Chung2020PRBPlaquette}.
The PDW wave vector is near $(0, \pi)$ and settles into specific commensurate discrete momenta depending on the cylinder geometry [c.f. Fig.~\ref{Fig:EM_Phiyy} and Table~\ref{Tab:Q_PDW}].
Our finite-temperature results retrospectively explain some previous observations. 
For example, PDW states have been stabilized through ensemble tuning~\cite{Chen2025PNASGlobal} or geometry variation~\cite{Xu2024Stripes} in DMRG simulations, and also identified as low‑lying competing states in iPEPS studies~\cite{Yue2024Pseudogap, Zheng2025CPCompeting}.

\begin{figure}[tbp]
\centering
\includegraphics[width = \linewidth]{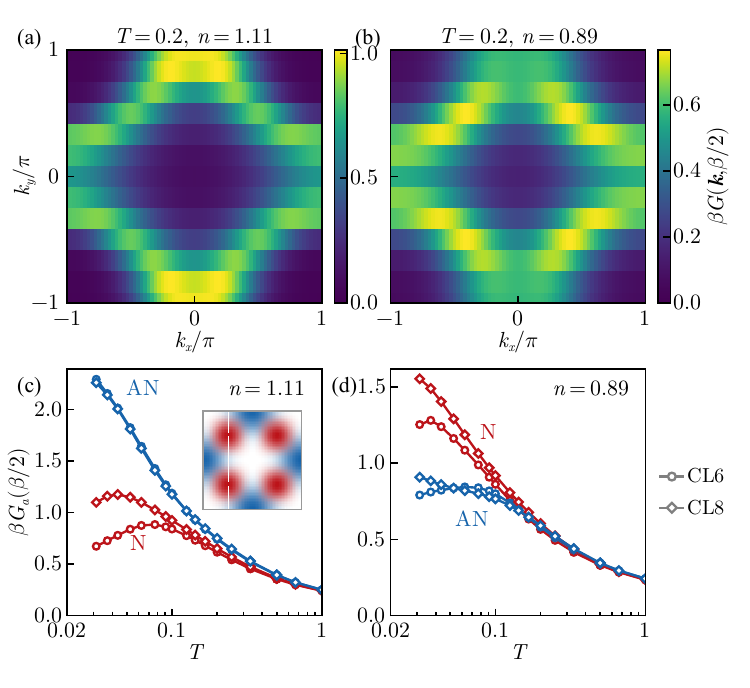}
\caption{(a,b) $\beta G(\mathbf{k}, \beta/2)$ in the first BZ at $T = 0.2$ for $1/9$ electron doping and hole doping, respectively. Combining PBC and APBC on a $6\times 18$ cylinder yields 12 discrete $k_y$ points. (c, d) Temperature evolution of $\beta G_a(\beta/2)$ with form factors $f_{\rm N}(\mathbf{k}) = 4(\sin k_x \cdot \sin k_y)^2$ (red) and $f_{\rm AN}(\mathbf{k}) = (\cos k_x - \cos k_y)^2$ (blue). These reweighting factors are illustrated in the inset of (c).}
\label{Fig:PG}
\end{figure}

\textcolor{blue}{\textit{Node-antinode dichotomy under doping.}}--- We demonstrate the particle-hole asymmetry in the electronic structure by computing $\beta G(\mathbf{k}, \beta/2)$ at fixed doping $\delta = \pm1/9$. As shown in Fig.~\ref{Fig:PG}(a,b), a profound evolution of the FS is observed at $T/t\simeq 0.2$, revealing distinct nodal-antinodal dichotomies for electron and hole doping.
For electron doping, the low-frequency spectral weight is predominantly concentrated near the antinodal regions [$\mathbf{k} = (0, \pm \pi)$ and $(\pm\pi, 0)$], despite the $C_4$ anisotropy introduced by the cylindrical geometry.
In contrast, under hole doping, the antinodal weight is severely suppressed, leaving the spectral weight concentrated in arc-like nodal regions [$\mathbf{k} = (\pm \pi/2, \pm \pi/2)$].
For clarity, we broadly refer to these arc-like features simply as ``Fermi arcs'', regardless of whether they strictly constitute unclosed FS segments.
Note such particle-hole asymmetric dichotomies are also observed in recent numerical studies~\cite{Qu2024PRLPhase, Wang2025Probing} and are consistent with angle‑resolved photoemission spectroscopy (ARPES) measurements on both hole‑ and electron‑doped cuprates~\cite{Matsui2007PRBEvolution, Armitage2010RoMPProgress, Shen2005SNodal}.

To quantify the node-antinode dichotomy, we define the reweighted single-particle Green's function as $G_a(\beta/2) = \frac{1}{N}\sum_{\mathbf{k}} f_a(\mathbf{k}) G(\mathbf{k}, \beta/2)$, where the form factors $f_{\rm N}(\mathbf{k}) = 4(\sin k_x \cdot \sin k_y)^2$ and $f_{\rm AN}(\mathbf{k}) = (\cos k_x - \cos k_y)^2$ select contributions from nodal and antinodal regions, respectively. We show the temperature dependence of $\beta G_a(\beta/2)$ in Fig.~\ref{Fig:PG}(c,d). A clear dichotomy emerges between the nodal and antinodal responses: under electron doping, the antinodal weight grows with temperature while the nodal part is suppressed. The opposite trend is observed for hole doping. This contrasting behavior is robust against variations in the cylinder geometry.

\begin{figure}[tbp]
\centering 
\includegraphics[width = \linewidth]{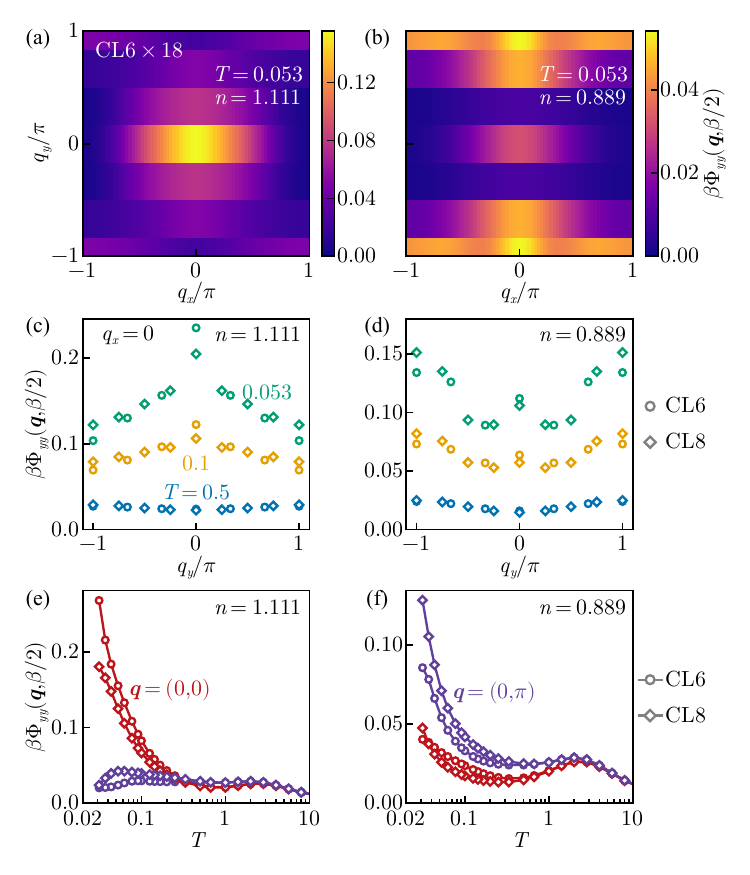}
\caption{(a,b) Contour plot of $\beta \Phi_{yy}(\mathbf{q}, \beta/2)$ averaged by PBC and APBC on a $6\times 18$ cylinder, at a representative temperature $T = 0.053$.
(c,d) $q_x = 0$ cut of $\beta \Phi_{yy}(\mathbf{q}, \beta/2)$ on $W\times 18$ cylinders with $W = 6, 8$, at several temperatures. Data at each fixed temperature are vertically offset by $0.04$ for the presentation.
(e,f) Temperature evolution of $\beta \Phi_{yy}(\mathbf{q}, \beta/2)$ at $\mathbf{q} = (0, 0)$ and $(0, \pi)$.
Data in (a,c,e) and (b,d,f) correspond to electron doping ($\delta = -1/9$) and hole doping ($\delta = 1/9$), respectively.}
\label{Fig:Phiyy}
\end{figure}

\textcolor{blue}{\textit{dSC versus PDW fluctuations.}}--- 
We then calculate $\beta \Phi_{yy}(\mathbf{q}, \beta/2)$ to examine the pairing instabilities. Figure~\ref{Fig:Phiyy}(a,b) displays contour plots of $\beta \Phi_{yy}(\mathbf{q}, \beta/2)$ at $T/t = 0.053$, a temperature at which pairing fluctuations dominate while CDW fluctuations remain weak. On the electron-doped side, the dominant pairing signal appears at $\mathbf{q} = (0, 0)$, corresponding to conventional zero-momentum dSC.
Conversely, on the hole-doped side, a pronounced peak emerges at $\mathbf{q} = (0, \pi)$, corresponding to dominant PDW fluctuations. This is clearly resolved by the ITP, whose kernel selects low‑energy fluctuations [Eq.~\eqref{Eq:ITP_Phi}]. In clear contrast, the equal-time structure factor $\Phi_{yy}(\mathbf{q})$ features a broad peak at $\mathbf{q} = 0$~\cite{Supplementary}, as it incorporates high-frequency and short-ranged fluctuations.
Moreover, in Fig.~\ref{Fig:Phiyy}(b) we observe another, weak pairing correlation near $\mathbf{q} = (\pi, \pi)$, which may be relevant to the $\eta$-pairing in Hubbard model~\cite{Yang1989PRLpairing, Moudgalya2020PRB}.

In Fig.~\ref{Fig:Phiyy}(c,d), we show the $q_x = 0$ cut of $\beta \Phi_{yy}(\mathbf{q}, \beta/2)$. At a relatively high temperature $T = 0.5$, the signal increases monotonically with $ 0 \leq |q_y| \leq \pi$ for both electron and hole doping. A notable difference occurs upon cooling for electron doping: around $T/t = 0.1$, the values decrease with $|q_y|$ --- a trend sharply contrasting with the hole-doped side at the same temperature. This qualitative difference between the two doping cases becomes even more pronounced at $T/t=0.053$. Although a zero-momentum pairing signature is also detectable under hole doping, it is significantly weaker than both the $(0, \pi)$-PDW signal at the same doping and the zero-momentum pairing observed under electron doping.

The temperature dependence of $\beta \Phi_{yy}(\mathbf{q}, \beta/2)$ at relevant momenta, i.e.  $\mathbf{q} = 0$ (related to dSC) and $\mathbf{Q}_{\rm PDW} = (0, \pi)$, are shown in Fig.~\ref{Fig:Phiyy}(e,f). On the electron-doped side, $\beta \Phi_{yy}(0, \beta/2)$ and $\beta \Phi_{yy}(\mathbf{Q}_{\rm PDW}, \beta/2)$ are small and comparable at high temperature. Upon cooling, however, the two responses deviate strongly. The zero‑momentum pairing structrue factor exhibits a clear divergence, whereas the PDW correlations remain weak. On the other hand, $\beta \Phi_{yy}(\mathbf{Q}_{\rm PDW}, \beta/2)$ shows a similar divergence, although it eventually gives way to CDW order in the ground state~\cite{Jiang2024PRBGround, Jiang2025Competition}. Thus, a fluctuating PDW phase exists in an intermediate temperature window.
Moreover, the relation $\beta \Phi_{yy}(\mathbf{Q}_{\rm PDW}, \beta/2) > \beta \Phi_{yy}(0, \beta/2)$ persists to a relatively high temperature where the width-dependence is negligible ($0.5 \lesssim T/t \lesssim 3$). This robustness against finite‑size effects suggests that the dominant PDW fluctuation is an intrinsic property of the 2-dimensional limit.

\begin{figure}[tbp]
\centering
\includegraphics[width = \linewidth]{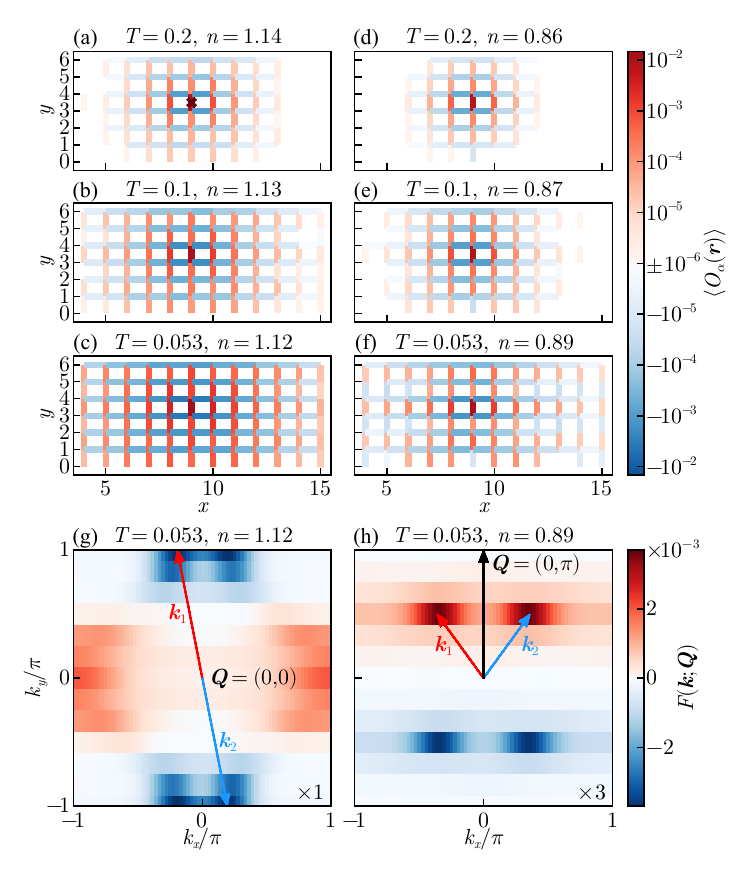}
\caption{(a-f) Real-space pairing $\expval{O_{\alpha}(\mathbf{r})}$ in the bulk of a CL$6\times 18$ cylinder. The local pairing field $h_\textrm{p} = 0.1$ is labeled by a cross mark in (a). The magnitude is shown with a logarithmic scale with a cutoff of $10^{-6}$. (g,h) Equal-time anomalous Green's function $F(\mathbf{k}; \mathbf{Q})$ with $\mathbf{Q} = (0, 0)$ for dSC and $\mathbf{Q} = (0, \pi)$ for PDW, respectively. The arrows illustrate the pairing momenta $\mathbf{k}_1$, $\mathbf{k}_2$ and the Cooper pair momentum $\mathbf{Q} = \mathbf{k}_1 + \mathbf{k}_2$. The chemical potential is fixed as $\mu = 6.1$ for electron doping and $\mu = 1.5$ for hole doping, with density $n$ indicated in each panel. The intensity in (h) is scaled up by a factor of $3$ for clarity.}
\label{Fig:RS}
\end{figure}

\textcolor{blue}{\textit{Pairing symmetries.}}--- To probe the static pairing susceptibility, we apply a local pairing field $-h_{\rm p}O_y(\mathbf{r}_0)$ with $h_p = 0.1$ on a central y-bond. This term explicitly breaks charge conservation and translation symmetry, enabling direct observation of pairing response at finite temperature.  
As shown in Fig.~\ref{Fig:RS}(a-c), pairing order $\expval{O_\alpha(\mathbf{r})}$ firstly emerges at the pinned bond and then spreads outward upon lowering temperature. 
For electron doping, applying the pairing field on a single $y$-bond induces responses of opposite sign on $x$-bonds, confirming the spontaneous emergence of $d$-wave pairing symmetry.
Under hole doping, however, a 2-periodic modulation along the $y$-direction with alternating signs develops, consistent with the $(0,\pi)$-PDW instability identified earlier.
Moreover, the short-ranged $d$-wave pattern remains visible, explaining the weak zero-momentum signal in Fig.~\ref{Fig:Phiyy}(b).

We also conduct an analysis of pairing symmetries in the momentum space, by computing the equal-time anomalous Green's function~\cite{Wang2025Anomalous}
\begin{equation}
     F(\mathbf{k}; \mathbf{Q}) = \frac{e^{i\theta}}{2}\left(\expval{
          c_{\mathbf{k}\downarrow}c_{\mathbf{Q} - \mathbf{k},\uparrow}} - \expval{c_{\mathbf{k}\uparrow}c_{\mathbf{Q} - \mathbf{k}\downarrow}}\right),
     \label{Eq:FkQ}
\end{equation}
which reflects the singlet pairing between electrons with momenta $\mathbf{k}$ and $\mathbf{Q} - \mathbf{k}$, where $\mathbf{Q}$ is the net momentum as well as the PDW vector. For $\mathbf{Q}\neq 0$, choice of the origin ($\mathbf{r} = 0$) leads to an arbitrary overall phase $\theta$. We set $\theta = \mathbf{Q}\cdot (\mathbf{r}_0+\hat{\mathbf{y}}/2)$ so that $F(\mathbf{k};\mathbf{Q})$ is real due to the mirror-$y$ symmetry with respect to the pinned bond.

As shown in Fig.~\ref{Fig:RS}(g), $F(\mathbf{k}; \mathbf{Q} = 0)$ at dSC phase exhibits a clear sign change along the $\pi/2$-rotation as expected, despite a distortion due to the cylindrical geometry. Moreover, the pairing occurs between the anti-nodal regions with higher single-particle spectral weight [see Fig.~\ref{Fig:PG}(a)], indicating the $d$-wave pairing near FS.
On the hole-doped side, we focus on the $(0, \pi)$-PDW and show $F(\mathbf{k}; \mathbf{Q} = (0, \pi))$ in Fig.~\ref{Fig:RS}(h). Four  peaks can be observed near the nodal points with $\mathbf{k} = (\pm \pi/2, \pm\pi/2)$, where the two peaks at $\mathbf{k}_1, \mathbf{k}_2$ contribute to the pairing with net momentum $\mathbf{Q} = (0, \pi)$.
The sign reversal under $y$-reflection shown in Fig.~\ref{Fig:RS}(h) agrees with the site-centered $p_y$-wave symmetry observed in the real-space pattern of Fig.~\ref{Fig:RS}(f). 
It is important to note that the $p$-wave symmetry, often linked to triplet pairing, is not a necessary condition for PDW order~\cite{Chen2020CPLSuperconductivity, Kuhlenkamp2025Robust}.

\textcolor{blue}{\textit{Discussion and outlooks.}}--- Whether the $t$-$t^\prime$ Hubbard model can fully capture high-$T_c$ superconductivity and the rich variety of electronic phases in the cuprate phase diagram remains an open and actively studied question~\cite{Jiang2024PRBGround, Xu2024SCoexistence, Zhang2025PRLFrustration}. 
Here, we present a finite-temperature study that characterizes SC pairing behaviors and maps out the temperature-doping phase diagram.
Using state-of-the-art tensor-network simulations, we find that the $d$-wave SC order is notably stronger under electron doping. In contrast, on the hole-doped side, a fluctuating PDW with wave vector $\mathbf{Q}_{\rm PDW} \approx (0, \pi)$ emerges at comparable temperatures, and gives way to CDW at lower temperature.
Therefore, the $t$-$t^\prime$ Hubbard model alone is insufficient to capture the rich phenomenology of hole-doped cuprates, suggesting that extensions beyond the single-band framework, such as density-assisted hopping~\cite{Jiang2023PRBDensity, Kovalska2025Tangent}, electron-phonon coupling~\cite{Wang2025SBRobust}, or the full three-band Emery model~\cite{Emery1987PRLTheory}, may be essential for a unified description.

Despite this, the Hubbard model provides an intriguing playground for investigating the interplay between PG and SC pairing~\cite{Emery1995NImportance, Preuss1997PRLPseudogaps, Gull2013PRLSuperconductivity, Lee2014PRXAmperean}. In the PG regime, spectral weights near the ``hot spots'' are suppressed by AFM fluctuations~\cite{Senechal2004PRLHot, Wang2025Probing}, resulting in residual anti-nodal (nodal) Fermi arcs in the electron-doped (hole-doped) regime. Regarding SC pairing, distinct electronic structures lead to particle-hole asymmetric pairing instabilities. Under electron doping, conventional opposite-momenta pairing at the anti-nodal regions yields a uniform dSC phase; on the hole-doped side, inter-arc pairing produces a PDW state with finite net momentum $\mathbf{Q}_{\rm PDW} \approx (0, \pi)$. The absence of Fermi-surface nesting in the latter case hinders condensation, consistent with the fluctuating nature of the PDW state.

Remarkably, we find that PDW fluctuations emerge in the lower portion of the underdoped PG regime. The underlying Fermi arc structure provides a natural explanation for this pairing instability, as it facilitates inter-arc pairing. The observed overlap between PDW and PG states in the temperature-doping phase diagram of $t$-$t'$ Hubbard model offers a fresh perspective on their relationship, opening new avenues to clarify their intertwined microscopic origins~\cite{Lee2014PRXAmperean, Fradkin2015RMPColloquium, Agterberg2020ARoCMPPhysics, Yue2024Pseudogap}.

\textit{Acknowledgments}.---Q.L. and W.L. are indebted to Dai-Wei Qu, Jialin Chen and Shou-Shu Gong for stimulating discussions. The tanTRG simulations are based on the open-source package FiniteMPS.jl~\cite{Li2026FiniteMPS.jl}.
This work is supported by the National Natural Science Foundation of China (Grant Nos.~12534009, and 12447101), and the Innovation Program for Quantum Science and Technology (Grand No. 2021ZD0301900). We thank HPC at ITP-CAS for the technical support and generous allocation of CPU time.

\bibliography{Hubbard.bib}

\clearpage
\newpage
\onecolumngrid
\begin{center}
\textbf{Appendix}
\end{center}
\twocolumngrid

\begin{figure}[htbp]
     \centering
     \includegraphics[width=\linewidth]{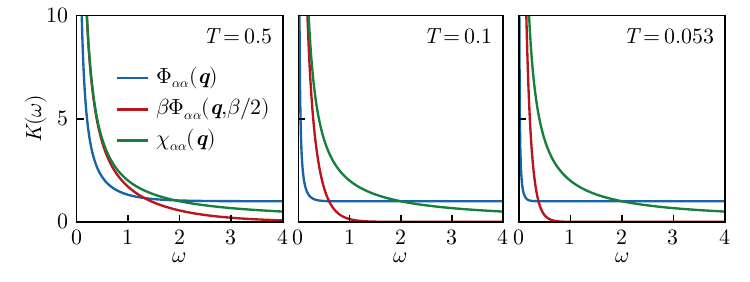}
     \caption{Comparison of the convolution kernels $1 / \tanh(\beta\omega/2)$, $\beta / \sinh(\beta\omega/2)$ and $2/\omega$ corresponding to the equal-, imaginary-time structure factors [$\Phi_{\alpha\alpha}(\mathbf{q})$ and $\beta \Phi_{\alpha\alpha}(\mathbf{q}, \beta/2)$] and static pairing susceptibility $\chi_{\alpha\alpha}(\bold{q})$, respectively, at representative temperatures $T = 0.5, 0.1$ and $0.053$.}
     \label{Fig:EM_kernel}
\end{figure}

\begin{figure}[htbp]
     \centering 
     \includegraphics[width = \linewidth]{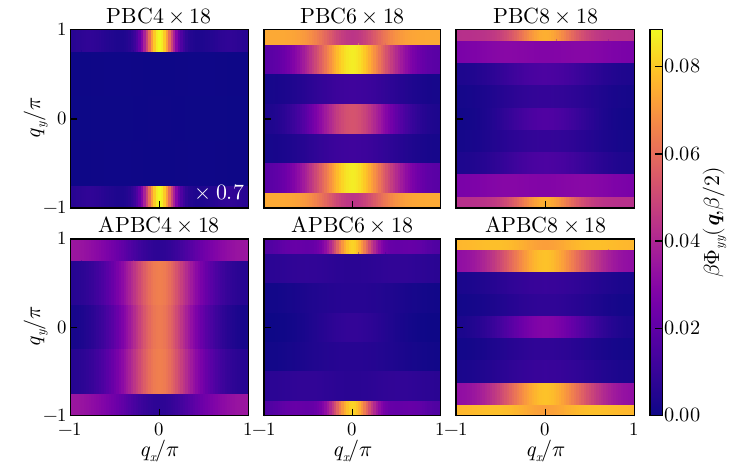}
     \caption{Geometry dependence of $\beta \Phi_{yy}(\mathbf{q}, \beta/2)$ at $1/9$ hole doping and $T = 0.053$. The data obtained on a PBC$4\times18$ cylinder is scaled by a factor of 0.7 for clarity.}
     \label{Fig:EM_Phiyy}
\end{figure}

\begin{figure}[htbp]
     \centering 
     \includegraphics[width = \linewidth]{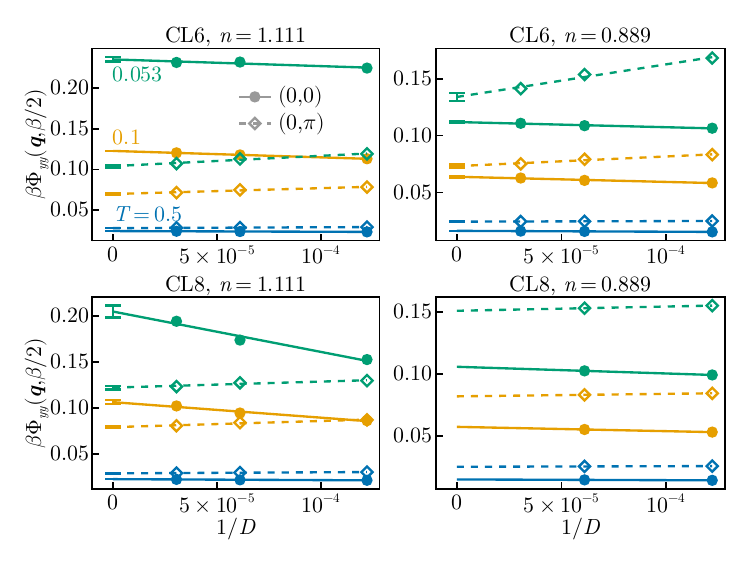}
     \caption{Extrapolation of $\beta\Phi_{yy}(\mathbf{q}, \beta/2)$ with respect to bond dimension $D$ at representative temperatures. The vertical offsets are identical to those in Fig.~\ref{Fig:Phiyy}(c,d).}
     \label{Fig:EM_extrap}
\end{figure}

\textcolor{blue}{\textit{Different quantities for probing the pairing instability.}}---
The dynamical pairing susceptibility is defined as 
\begin{equation}
     \chi_{\alpha\beta}(\mathbf{q}, \omega) = i \int_{0}^\infty dt e^{i(\omega+i0^+) t}\expval{\left[O_\mathbf{q}^\alpha(t), O_\mathbf{-q}^\beta\right]},
\end{equation} 
where $O_\mathbf{q}^\alpha$ is the pairing operators defined previously. Focusing on the diagonal component $\alpha = \beta$, the equal-, imaginary-time pairing structure factors and static pairing susceptibility follow the spectral representations as
\begin{equation}
     \left\{
          \begin{aligned}
               &\Phi_{\alpha\alpha}(\mathbf{q}) = \int_{0}^\infty \frac{d\omega}{\pi}  \frac{\Im\chi_{\alpha\alpha}(\mathbf{q}, \omega)}{\tanh(\beta\omega/2)},\\
               &\beta\Phi_{\alpha\alpha}(\mathbf{q}, \beta/2) = \int_{0}^\infty \frac{d\omega}{\pi}  \frac{\beta\Im\chi_{\alpha\alpha}(\mathbf{q}, \omega)}{\sinh(\beta\omega/2)},\\
               &\chi_{\alpha\alpha}(\mathbf{q}) = \int_{0}^\infty \frac{d\omega}{\pi}  \frac{\Im\chi_{\alpha\alpha}(\mathbf{q}, \omega)}{\omega/2},
          \end{aligned}
     \right.
     \label{Eq:SR_all}
\end{equation}  
proved via Lehmann representations in Supplemental Material~\cite{Supplementary}.
At lower temperatures, $\Phi_{\alpha\alpha}(\bold{q}, \beta/2)$ and $\chi_{\alpha\alpha}(\bold{q})$ become increasingly sensitive to low-frequency pairing fluctuations, as the convolution kernels $1/\sinh(\beta\omega/2) \sim e^{-\beta\omega/2}$ and $2/\omega$ suppress high-frequency contributions exponentially and algebraically, respectively, as shown in Fig.~\ref{Fig:EM_kernel}.

\textcolor{blue}{\textit{PDW wave vectors.}}---
To illustrate the geometry dependence, we focus on $1/9$ hole doping at $T = 0.053$ and present the separate results of $\beta\Phi_{yy}(\mathbf{q}, \beta/2)$ in Fig.~\ref{Fig:EM_Phiyy} for PBC and APBC on cylinders with $W = 4, 6$ and $8$.
Across all simulated geometries, the strongest pairing tendency shifts to a non-zero momentum, although the specific value varies, as summarized in Table~\ref{Tab:Q_PDW}.
Specifically, we find the PDW momentum occurs at $q_y = \pi$ as long as $k_y = \pi/2$ is an available single-particle momentum. Otherwise, the system favors the nearest available momentum, e.g., $q_y = 3\pi/4$ for $W = 8$ with APBC.

\begin{table}[htbp]
     \centering
     \caption{Geometry dependence of PDW wave vector $\mathbf{Q}_{\rm PDW}$ at $1/9$ hole doping.}

     \begin{tabularx}{\linewidth}{|Y|Y|Y|Y|Y|Y|}
          \hline
          \multicolumn{2}{|c|}{$W = 4$} & \multicolumn{2}{c|}{$W = 6$} & \multicolumn{2}{c|}{$W = 8$} \\
          \hline
          PBC & APBC & PBC & APBC & PBC & APBC \\
          \hline
          $(0, \pi)$ & $(0, \pi/2)$ & $(0, 2\pi/3)$ & $(0, \pi)$ & $(0, \pi)$ & $(0, 3\pi/4)$ \\
          \hline
     \end{tabularx}
     \label{Tab:Q_PDW}
\end{table}

\textcolor{blue}{\textit{Extrapolation with bond dimensions.}}---
In Fig.~\ref{Fig:EM_extrap}, we represent the extrapolation of $\beta \Phi_{yy}(\mathbf{q}, \beta/2)$ to the infinite bond dimension limit ($D \rightarrow \infty$), at parameters relevant to Fig.~\ref{Fig:Phiyy}.
We find increasing $D$ consistently enhances the zero-momentum pairing [$\beta \Phi_{yy}(0, \beta/2)$], which is a common behavior in DMRG studies of a dSC state~\cite{Gong2021PRLRobust,Lu2024PRLEmergent}.
Conversely, the PDW fluctuations at $\mathbf{q} = (0, \pi)$ tend to be overestimated at smaller bond dimensions.
Nevertheless, a substantial gap persists between the extrapolated values at $\mathbf{q} = (0, \pi)$ and $\mathbf{q} = (0, 0)$ on the hole-doped side.
This confirms that while the finite bond dimension quantitatively shifts the relevant temperature scales, it does not alter the qualitative conclusion regarding the existence of the fluctuating PDW phase.

\clearpage
\newpage
\onecolumngrid
\begin{center}
\textbf{Supplemental Material for \\ \thetitle}
\end{center}

\setcounter{equation}{0}
\setcounter{figure}{0}
\setcounter{table}{0}
\setcounter{page}{1}
\makeatletter
\renewcommand{\thetable}{S\arabic{table}}
\renewcommand{\theequation}{S\arabic{equation}}
\renewcommand{\thefigure}{S\arabic{figure}}
\setcounter{secnumdepth}{3}

\section{Supplementary derivation of formulas in the main text}
\subsection{Imaginary-time proxy for pairing structure factor}
According to the linear response theory, the dynamical pairing susceptibility is defined as
\begin{equation}
     \chi_{\alpha\beta}(\bold{q}, \omega) = i \int_{0}^\infty dt e^{i(\omega+i0^+) t}\expval{\left[O_\bold{q}^\alpha(t), O_\bold{-q}^\beta\right]},
\end{equation} 
where $O_\bold{q}^\alpha$ is the pairing operator defined in the main text.
An alternative dynamical quantity characterizing pairing fluctuations is the dynamical pairing structure factor
\begin{equation}
     \Phi_{\alpha\beta}(\bold{q}, \omega) = \int_{-\infty}^\infty dt e^{i\omega t}\expval{O_\bold{q}^\alpha(t)O_\bold{-q}^\beta}.
\end{equation} 
Using the property $\left(O_\bold{q}^{\alpha}\right)^\dagger = O_{-\bold{q}}^\alpha$, the Lehmann representations for the diagonal components ($\alpha = \beta$) are given by
\begin{equation}
    \Im\chi_{\alpha\alpha}(\bold{q}, \omega) = \frac{\pi}{\mathcal{Z}}\sum_{mn}|\bra{m}O_\bold{q}^\alpha\ket{n}|^2(e^{-\beta E_m} - e^{-\beta E_n})\delta(\omega + E_m - E_n),
    \label{Eq:Lehmann_chi}
\end{equation}
and 
\begin{equation}
     \Phi_{\alpha\alpha}(\bold{q}, \omega) = \frac{2\pi}{\mathcal{Z}}\sum_{mn}|\bra{m}O_\bold{q}^\alpha\ket{n}|^2e^{-\beta E_m}\delta(\omega + E_m - E_n),
     \label{Eq:Lehmann_Phi}
\end{equation}
where $\ket{n}$ denotes the eigenstate with energy $E_n$, and $\mathcal{Z}$ is the partition function. 
Combining Eq.~(\ref{Eq:Lehmann_chi}) and Eq.~(\ref{Eq:Lehmann_Phi}) yields the fluctuation-dissipation theorem $2{\rm Im}\chi_{\alpha\alpha}(\bold{q}, \omega) = (1 - e^{-\beta\omega})\Phi_{\alpha\alpha}(\bold{q}, \omega)$. 
Recognizing that ${\rm Im}\chi_{\alpha\alpha}(\bold{q}, \omega)$ is proportional to the bosonic spectral function, we obtain the spectral representation
\begin{equation}
     \Phi_{\alpha\alpha}(\bold{q}, \beta/2) = \int_{-\infty}^\infty \frac{d\omega}{2\pi}e^{-\beta\omega/2}\Phi_{\alpha\alpha}(\bold{q}, \omega) = \int_{-\infty}^\infty \frac{d\omega}{2\pi}\frac{{\rm Im}\chi_{\alpha\alpha}(\bold{q}, \omega)}{\sinh(\beta\omega/2)} = \frac{\pi}{\beta}\int_{-\infty}^\infty d\omega K(\omega; \beta)\frac{\Im\chi_{\alpha\alpha}(\bold{q}, \omega)}{\beta\omega},
     \label{Eq:ITP_chi}
\end{equation}
where the kernel function $K(\omega; \beta) = \beta^2\omega/[2\pi^2\sinh(\beta\omega/2)]$ peaks at $\omega = 0$ and approaches $\delta(\omega)$ in the limit $\beta \rightarrow \infty$.
Consequently, Eq.~(\ref{Eq:ITP_chi}) justifies the imaginary-time proxy (ITP)  
\begin{equation}
     \Phi_{\alpha\alpha}(\bold{q}, \omega = 0) = \lim_{\omega \rightarrow 0}\frac{2{\rm Im}\chi_{\alpha\alpha}(\bold{q}, \omega)}{\beta\omega} \approx \frac{2\beta}{\pi}\Phi_{\alpha\alpha}(\bold{q}, \beta/2)   
\end{equation}
in the low-temperature limit.
Even at relatively high temperatures, this quantity remains an indicator of low-frequency spectral weight ($\omega \lesssim 1/\beta$) due to the single-peaked nature of the kernel. 

By exchanging indices $m$ and $n$ in Eq.~(\ref{Eq:Lehmann_chi}), we find $\Im\chi_{\alpha\alpha}(\bold{q}, -\omega) = -\Im\chi_{\alpha\alpha}(-\bold{q}, \omega)$. Further considering the spatial reflection symmetry, $\Im \chi_{\alpha\alpha}(\bold{q}, \omega)$ becomes an odd function of $\omega$. Therefore, the spectral representation of the equal-time structure factor is simplified as 
\begin{equation}
     \Phi_{\alpha\alpha}(\bold{q}) \equiv \expval{O_\bold{q}^\alpha O_\bold{-q}^\alpha} = \int_{-\infty}^{\infty} \frac{d\omega}{\pi} \frac{\Im \chi_{\alpha\alpha}(\bold{q}, \omega)}{1 - e^{-\beta\omega}} = \int_{0}^\infty \frac{d\omega}{\pi}  \frac{\Im\chi_{\alpha\alpha}(\bold{q}, \omega)}{\tanh(\beta\omega/2)}.
     \label{Eq:SR_Phi}
\end{equation}
Similarly, Eq.~\eqref{Eq:ITP_chi} can be rewritten as
\begin{equation}
     \Phi_{\alpha\alpha}(\bold{q}, \beta/2) \equiv \expval{e^{\beta\omega/2}O_\bold{q}e^{-\beta\omega/2}O_\bold{-q}} = \int_{-\infty}^{\infty} \frac{d\omega}{\pi} \frac{e^{-\beta\omega/2} \Im \chi_{\alpha\alpha}(\bold{q}, \omega)}{1 - e^{-\beta\omega}} = \int_{0}^\infty \frac{d\omega}{\pi}  \frac{\Im\chi_{\alpha\alpha}(\bold{q}, \omega)}{\sinh(\beta\omega/2)}.
     \label{Eq:SR_Phi_ITP}
\end{equation} 
Furthermore, the static susceptibility is related to $\Im\chi_{\alpha\alpha}(\bold{q},\omega)$ via the Kramers-Kronig relation as
\begin{equation}
     \chi_{\alpha\alpha}(\bold{q}) \equiv \chi_{\alpha\alpha}(\bold{q}, \omega \rightarrow 0) = \int_{-\infty}^{\infty} \frac{d\omega}{\pi} \frac{\Im \chi_{\alpha\alpha}(\bold{q}, \omega)}{\omega} = \int_{0}^\infty \frac{d\omega}{\pi} \frac{\Im \chi_{\alpha\alpha}(\bold{q}, \omega)}{\omega/2}.
\end{equation}

\subsection{Imaginary-time proxy for single-particle spectral function}
The Lehmann representation of single-particle Green's function is
\begin{equation}
     A(\mathbf{k}, \omega) = \frac{\pi}{\mathcal{Z}}\sum_{mn}(e^{-\beta E_m} + e^{-\beta E_n})|\bra{m}c_{\mathbf{k}\sigma}\ket{n}|^2\delta(\omega + E_m - E_n),  
     \label{Eq:Lehmann_A}
\end{equation}
and the corresponding imaginary-time Green's function is
\begin{equation}
     G(\bold{k}, \beta/2) = \frac{1}{\mathcal{Z}}\sum_{mn} e^{-\beta (E_m + E_n)/2}|\bra{m}c_{\mathbf{k}\sigma}\ket{n}|^2.
     \label{Eq:Lehmann_ITP}
\end{equation}
Considering the spin SU(2) symmetry, the single-particle Green's function is independent of spin index $\sigma$.
Comparing Eq.~(\ref{Eq:Lehmann_A}) and Eq.~(\ref{Eq:Lehmann_ITP}) yields the spectral representation 
\begin{equation}
     G(\bold{k}, \beta/2) = \int_{-\infty}^{\infty} \frac{d\omega}{2\pi} \frac{A(\bold{k}, \omega)}{2\cosh(\beta\omega/2)} = \frac{1}{2\beta}\int_{-\infty}^{\infty} d\omega K(\omega;\beta) A(\bold{k}, \omega), 
\end{equation}
where the kernel function $K(\omega;\beta) = \beta/[2\pi\cosh(\beta\omega/2)]$ peaks at $\omega = 0$ and approaches $\delta(\omega)$ as $\beta \rightarrow \infty$. This justifies the imaginary-time proxy $A(\bold{k},\omega = 0) \approx 2\beta G(\bold{k}, \beta/2)$ used in the main text.

\subsection{Form factors in reweighted Green's functions}
To distinguish contributions from different regions of the first Brillouin zone (BZ), we define the reweighted Green's functions using specific form factors as 
\begin{equation}
     G_a(\beta/2) = \frac{1}{N}\sum_{\mathbf{k}} f_a(\mathbf{k}) G(\bold{k}, \beta/2),
\end{equation} 
where $a$ denotes the type of form factor. In this work, we employ $f_{\rm N}(\mathbf{k}) = 4(\sin k_x \cdot \sin k_y)^2$ and $f_{\rm AN}(\mathbf{k}) = (\cos k_x - \cos k_y)^2$ to extract nodal and anti-nodal contributions, respectively. The corresponding real-space formulations are 
\begin{equation}
     \left\{
     \begin{aligned}
          & G_{\rm N}(\beta/2) = \expval{c_{i\sigma}(\beta/2)c_{i\sigma}^\dagger} - \frac{1}{2}\sum_{r_{ij} = 2}\expval{c_{i\sigma}(\beta/2)c_{j\sigma}^\dagger} + \frac{1}{4} \sum_{r_{ij} = 2\sqrt{2}} \expval{c_{i\sigma}(\beta/2)c_{j\sigma}^\dagger},\\
          & G_{\rm AN}(\beta/2) = \expval{c_{i\sigma}(\beta/2)c_{i\sigma}^\dagger} - \frac{1}{2}\sum_{r_{ij} = \sqrt{2}}\expval{c_{i\sigma}(\beta/2)c_{j\sigma}^\dagger} + \frac{1}{4} \sum_{r_{ij} = 2} \expval{c_{i\sigma}(\beta/2)c_{j\sigma}^\dagger},\\
     \end{aligned}
     \right.
\end{equation}
where $i$ is the reference site.

\section{Data convergence}
As a MPO-based method, the accuracy of tanTRG is controlled by the bond dimension $D$, analogous to the bond dimension of MPS in DMRG. However, while the low-temperature density operator may be highly entangled, the infinitely high-temperature density operator is a product state that can be exactly represented by a MPO with bond dimension $D = 1$, making the numerical accuracy of tanTRG highly related to the temperature. 
Therefore, comparing the same quantity obtained with different $D$ provides a posterior way to verify the reliability of the data at a given temperature. 
To support the findings in the main text, we show the data convergence with respect to $D$ of relevant quantities in this section.

\begin{figure*}[htbp]
     \centering 
     \includegraphics[width = \linewidth]{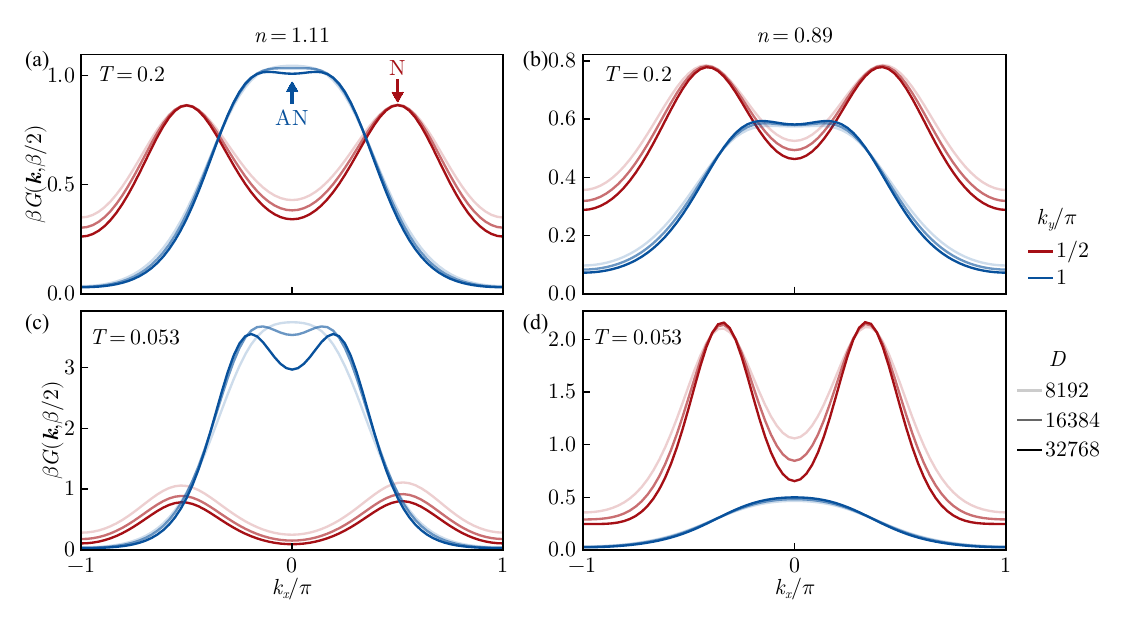}
     \caption{Convergence of $\beta G(\bold{k}, \beta/2)$ with respect to bond dimension $D$ at representative temperatures $T = 0.2$ (top) and $T = 0.053$ (bottom) for $1/9$ electron doping (left) and $1/9$ hole doping (right). The arrows in (a) highlight the nodal (N) and antinodal (AN) points.}
     \label{Fig:SM_PG_convergence}
\end{figure*}

In Fig.~\ref{Fig:SM_PG_convergence}, we show the convergence of $\beta G(\bold{k}, \beta/2)$ at temperatures $T = 0.2$ and $T = 0.053$ for both $1/9$ electron and hole doping, corresponding to Fig.~\ref{Fig:PG} in the main text. We focus on  $k_y = \pi/2$ and $k_y = \pi$, which cross nodal and antinodal points, respectively. Although the values at certain momenta persist to vary up to the largest $D = 32768$ we keep, the node-antinode features are qualitatively unchanged.
More specifically, we find that the finite-$D$ effect can be mainly characterized as broadening of peaks. This observation can be understood as the peaks in $\beta G(\bold{k}, \beta/2)$ reflects the FS or single-particle gap minima, while either of them leads to high entanglement, thus requires larger bond dimensions to capture.

\begin{figure*}[htbp]
     \centering 
     \includegraphics[width = \linewidth]{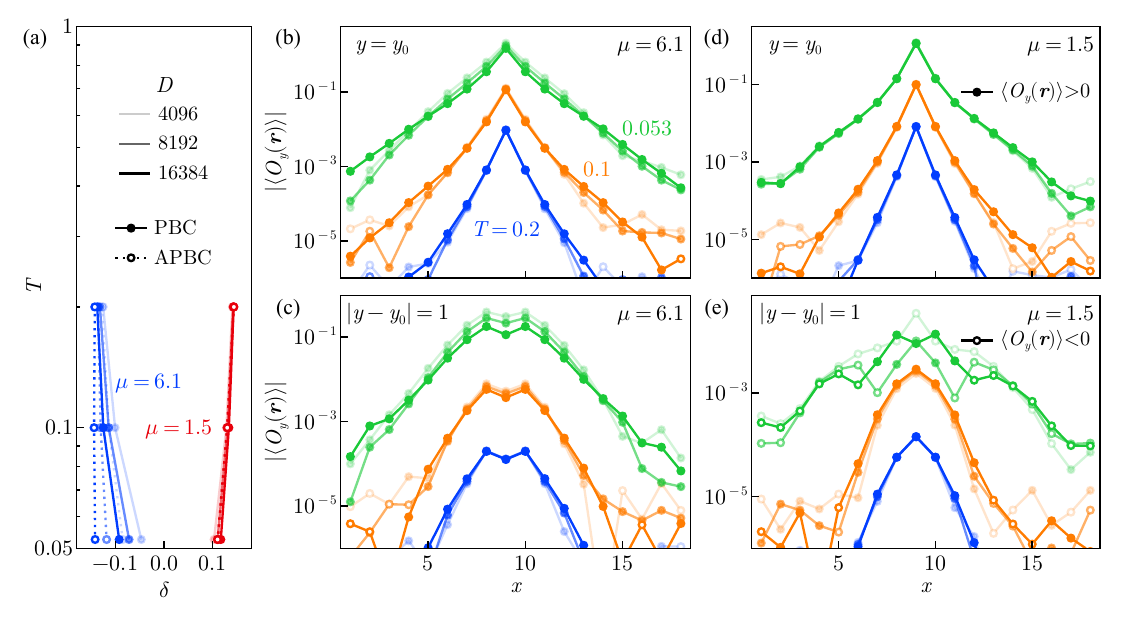}
     \caption{(a) Doping $\delta$ vs. temperature $T$ with fixed chemical potential $\mu = 6.1$ and $1.5$. (b,c) Local pairing order $\expval{O_y(\mathbf{r})}$ along the $x$-direction with fixed $y = y_0$ (b) and $|y - y_0| = 1$ (c) at electron doping ($\mu = 6.1$). (d,e) Similar to (b,c) but for hole doping ($\mu = 1.5$). The positive and negative data in (b-d) are marked as filled and empty circles, respectively. 
     All data are obtained on a six-leg cylinder, with a local pinning field $h_p = 0.1$ on the central $y$-bond at $\mathbf{r_0} = (9, 3)$.}
     \label{Fig:SM_Z2_convergence}
\end{figure*}

When there exists a nonzero pairing field that breaks the charge conservation symmetry, we cannot adopt the recently developed dynamical chemical potential technique to precisely control the doping level~\cite{Li2026PRBThermal}. Instead, we use fixed chemical potential $\mu = 6.1$ and $1.5$ for electron and hole doping, respectively, to characterize the pairing response [Fig.~\ref{Fig:RS} in the main text]. This makes the convergence with respect to $D$ more challenging, as the finite-$D$ effect constitutes not only the intrinsic error due to the representation ability of the MPO but also the deviation of the doping level. The latter effect is more significant at electron doping [Fig.~\ref{Fig:SM_Z2_convergence}(a)], which may be attributed to the phase separation phase zero temperature~\cite{Jiang2024PRBGround}. As a consequence, we can only obtain qualitatively converged results, as shown in Fig.~\ref{Fig:SM_Z2_convergence}(b-d). At higher temperatures [e.g. $T \geq 0.1$], the pairing response around the pairing field is well converged, as it relates to the short-ranged pairing susceptibility. At low temperatures [e.g. $T = 0.053$], the overall pairing response varies with $D$, which can be understood as the pairing response is highly sensitive to the doping level [c.f. Fig.~\ref{Fig:SM_Z2_convergence}(a)]. However, all results with different $D$ show the same qualitative behavior, i.e., exponentially decaying pairing response staring from the source position $x_0 = 9$. In particular, the sign change of $\expval{O_y(\mathbf{r})}$ versus $x$ with $|y - y_0|=1$ at hole doping is observed if $D \geq 8192$, which corresponds to the crossover between short-range dSC and $(0, \pi)$-PDW with higher correlation length.

\section{Supplementary numerical results}
\subsection{Real-space charge distribution}
\begin{figure*}[htbp]
     \centering 
     \includegraphics[width = \linewidth]{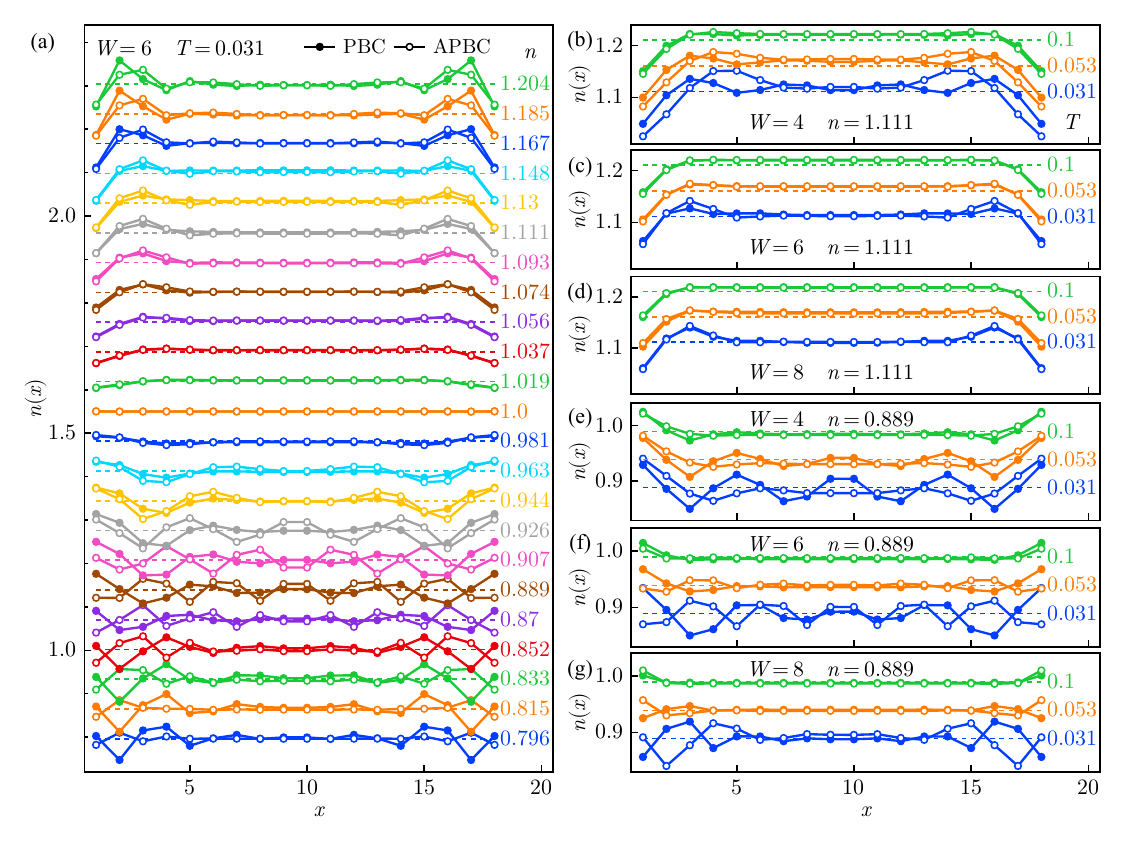}
     \caption{Real-space charge distribution $n(x)$. (a) Doping dependence at fixed cylinder width $W = 6$ and temperature $T = 0.031$. (b-d) Temperature dependence at $1/9$ electron doping, with $W = 4, 6$ and $8$, respectively. (e-g) Same as (b-d) but for $1/9$ hole doping. Curve are vertically offset by increments of 0.05 for clarity, with dashed lines of corresponding colors indicating mean values. 
     The bond dimensions are $D=8192$ for (a) and $D=16384$ for (b-g).}
     \label{Fig:SM_CDW}
\end{figure*}

{In Fig.~\ref{Fig:SM_CDW}(a), we represent the real-space charge distribution $n(x) = \frac{1}{W}\sum_{y=1}^W n(x,y)$ at a characteristic temperature ($T = 0.031$) of charge density wave (CDW)~\cite{Wietek2021PRXStripes, Li2026PRBThermal, Sinha2026Evolution} under varying doping up to $|\delta| \approx 0.2$.} On the electron-doped side, the charge distribution remains nearly uniform, except for short-ranged boundary oscillations extending fewer than approximately six sites. 
In contrast, on the hole-doped side, the charge modulation is systematically stronger than its electron-doped counterpart, particularly near $\delta = 0.1$.

Fig.~\ref{Fig:SM_CDW}(b-g) present the temperature and width dependence at $\delta = \pm 1/9$.
For $W = 6$, the CDW is virtually absent in the bulk for both doping sides at $T \gtrsim 0.053$. This confirms that $T = 0.053$ is a suitable representative temperature for analyzing pairing properties in the main text.
Furthermore, we observe that CDW orders are systematically suppressed as the cylinder width increases. 
For instance, at $1/9$ hole doping and $T = 0.053$, the charge distribution is nearly uniform for $W = 6$ and $W = 8$, while a clear modulation persists to the center of the $W = 4$ cylinder. 
This monotonic suppression with increasing $W$ is consistent with previous studies on $t$-$t^\prime$-$J$ model~\cite{Qu2024PRLPhase,Lu2024PRLEmergent}.

\subsection{Additional results of the single-particle spectral weight}
In this section, we provide supplementary results for $\beta G(\mathbf{k}, \beta/2)$ to further substantiate the distinct node-antinode structures between electron and hole doping reported in the main text.

\begin{figure*}[htbp]
     \includegraphics[width = \linewidth]{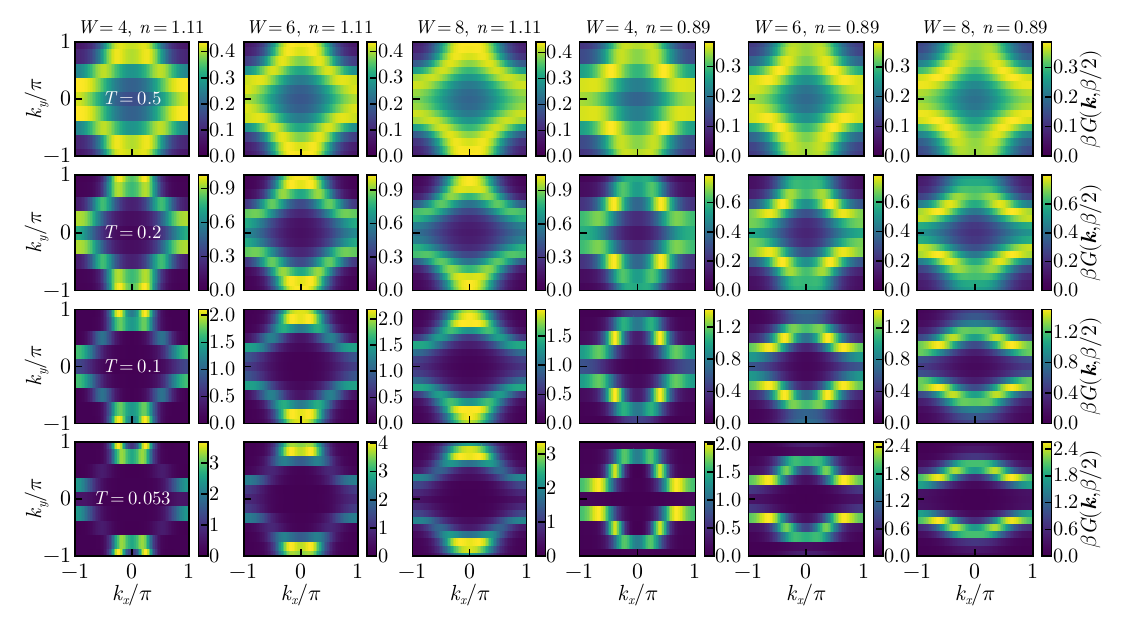}
     \caption{Contour plot of $\beta G(\bold{k}, \beta/2)$ in the first BZ. From top to bottom, rows correspond to temperatures $T = 0.5$, $0.2$, $0.1$ and $0.053$. Columns correspond to fixed cylinder widths $W$ and fillings $n$, as labeled in the top row. All data are obtained with bond dimension $D = 16384$.}
     \label{Fig:SM_G}
\end{figure*}

Fig.~\ref{Fig:SM_G} presents the cylinder-width dependence of $\beta G(\bold{k}, \beta/2)$ at dopings $\delta = \pm 1/9$ and temperatures $T = 0.5$, $0.2$, $0.1$ and $0.053$. At relatively high temperatures (e.g., $T = 0.5$), a hole-like Fermi surface (FS) with approximate $C_4$ symmetry is observed across all cylinder widths and for both doping sides.
As the temperature decreases and the correlation length increases, $C_4$ symmetry is broken by the cylindrical geometry. The finite bond dimension $D$ further introduces additional unphysical distortions [c.f. Fig.~\ref{Fig:SM_PG_convergence}], especially on wider cylinders.
We therefore select $W = 6$ and $T = 0.2$ as representative parameters for the main text, as they offer an optimal balance between data quality and the clarity of the node-antinode structure.  

\begin{figure*}[htbp]
     \includegraphics[width = \linewidth]{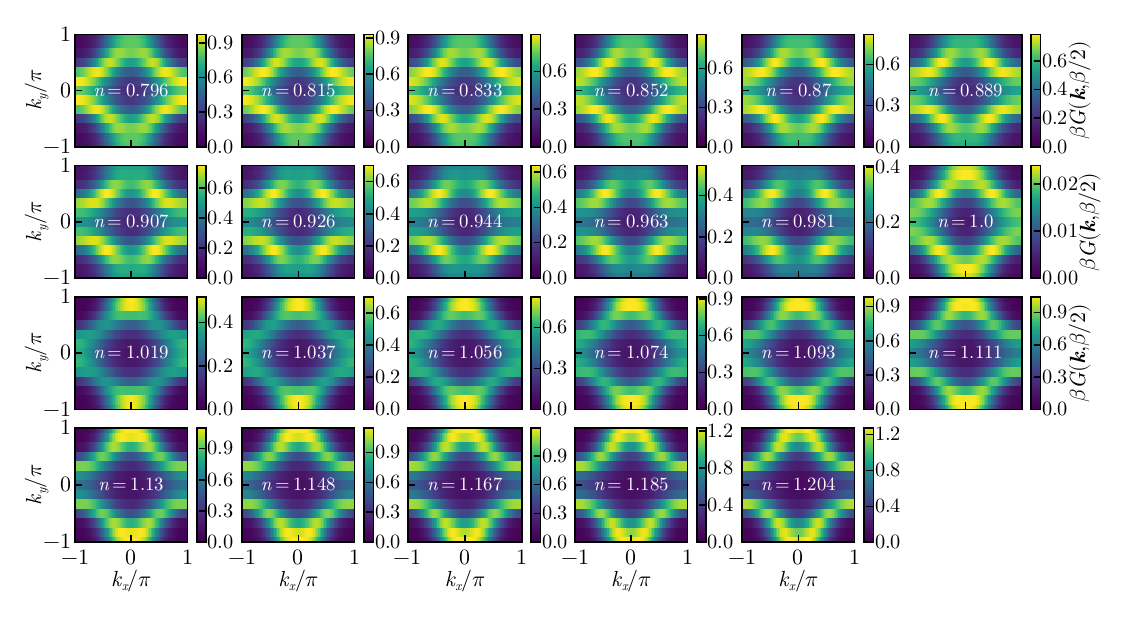}
     \caption{Doping dependence of $\beta G(\bold{k}, \beta/2)$ at fixed cylinder width $W = 6$ and temperature $T = 0.2$. All data are obtained with bond dimension $D = 8192$.}    
     \label{Fig:SM_G_doping} 
\end{figure*}

Next, we examine the doping dependence at fixed $W =6$ and $T = 0.2$, as shown in Fig.~\ref{Fig:SM_G_doping}. At half filling, the system is a Mott insulator, consistent with the significant suppression of $\beta G(\bold{k}, \beta/2)$ compared to the doped cases.
Across all investigated doping levels ($|\delta| \lesssim 0.2$), the anti-nodal [restricted to $(0, \pm\pi)$ due to distortion] and nodal regimes exhibit enhanced $\beta G(\bold{k}, \beta/2)$ values for electron and hole doping, respectively, consistent with the findings in the main text. This node-antinode dichotomy is more pronounced at lower doping levels, which can be understood as the FS is expected to become more homogeneous when approaching the anticipated Fermi-liquid regime at higher dopings.

\begin{figure}[htbp]
     \includegraphics[width = 0.55\linewidth]{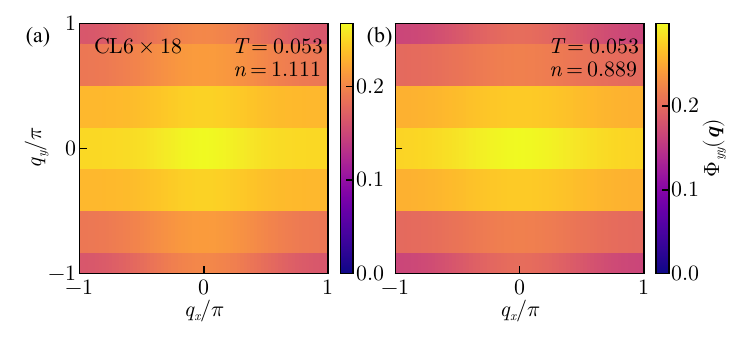}
     \caption{Equal-time pairing structure factor $\Phi_{yy}(\mathbf{q})$ under $1/9$ electron (a) and hole (b) doping. The temperature and geometry are fixed as $T = 0.053$ and a $6\times 18$ cylinder, respectively.}    
     \label{Fig:SM_Phiyy_EqualTime} 
\end{figure}

\subsection{Additional results of pairing structure factors}
We present the equal-time pairing structure factor $\Phi_{yy}(\mathbf{q})$ in Fig.~\ref{Fig:SM_Phiyy_EqualTime}, as a comparison to the imaginary-time correlations [c.f. Fig.~\ref{Fig:Phiyy}(a,b) in the main text], and find only a broadened peak at $\mathbf{q} = 0$ on both doping sides. It thus suggests that contributions from high frequency (and short-range correlations) dominate in $\Phi_{yy}(\mathbf{q})$.

\begin{figure*}[htbp]
     \includegraphics[width = \linewidth]{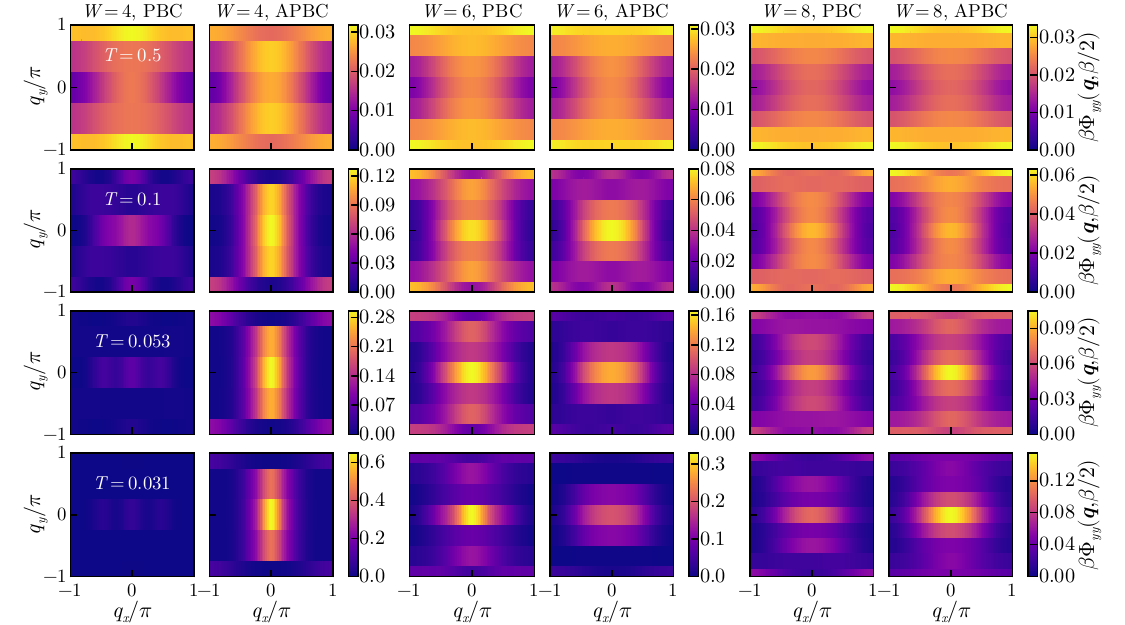}
     \caption{Geometry dependence of $\beta \Phi_{yy}(\bold{q}, \beta/2)$. From top to bottom, rows correspond to fixed representative temperatures $T = 0.5$, $0.1$, $0.053$ and $0.031$. Columns represent different geometries, specified by cylinder width $W$ and boundary conditions along the $y$ direction. All data are obtained at $1/9$ electron doping with fixed bond dimension $D = 16,384$.}
     \label{Fig:SM_Phi_e}
\end{figure*}

\begin{figure*}[htbp]
     \includegraphics[width = \linewidth]{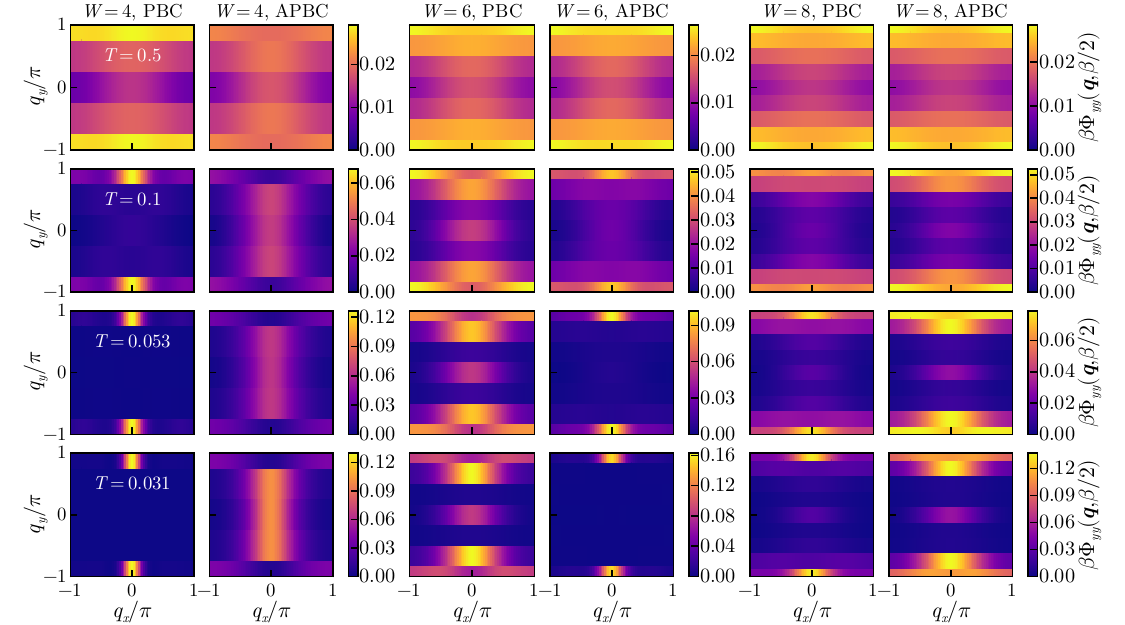}
     \caption{The $\beta \Phi_{yy}(\bold{q}, \beta/2)$ results, with the layout similar to Fig.~\ref{Fig:SM_Phi_e} but for $1/9$ hole doping.}
     \label{Fig:SM_Phi_h}
\end{figure*}

We represent the geometry dependence of $\beta \Phi_{yy}(\bold{q}, \beta/2)$ at $1/9$ electron doping in Fig.~\ref{Fig:SM_Phi_e}.
We find $\mathbf{q} = 0$ consistently remains the dominant pairing momentum at low temperatures, confirming the robust dSC on electron-doped side. Moreover, the difference between PBC and APBC decreases monotonically with cylinder width $W$, as expected since both boundary conditions should yield identical results in the 2D limit.
The hole-doped counterpart is shown in Fig.~\ref{Fig:SM_Phi_h}, where a clear signature of PDW is observed, with corresponding net momenta summarized in Table.~\ref{Tab:Q_PDW}
of the main text.

\begin{figure*}[htbp]
     \includegraphics[width = \linewidth]{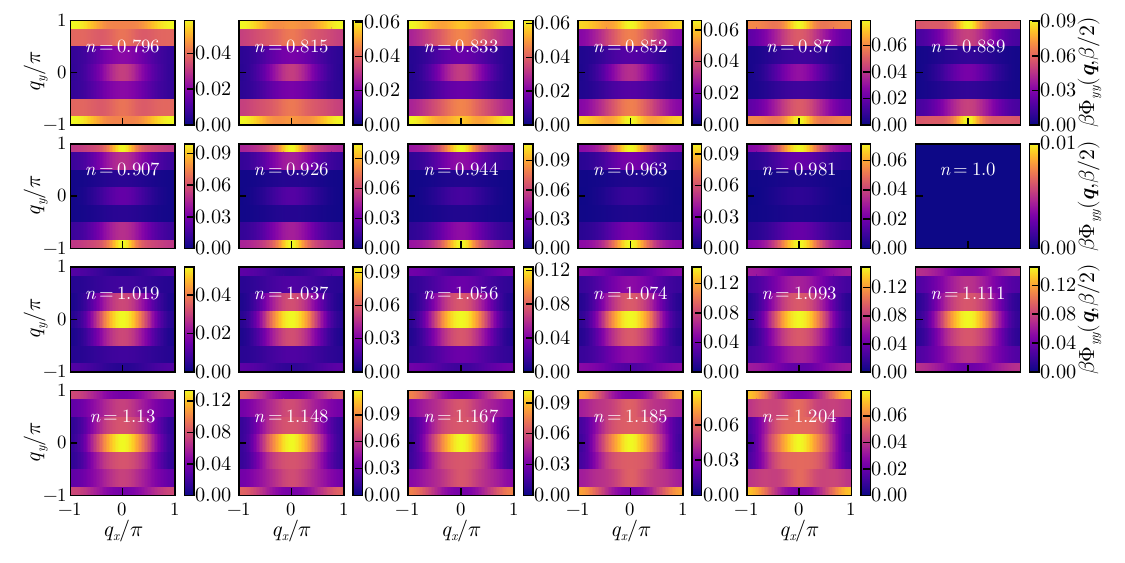}
     \caption{Doping dependence of $\beta \Phi_{yy}(\bold{q}, \beta/2)$ at fixed cylinder width $W = 6$ and temperature $T = 0.053$. All data are obtained with bond dimension $D = 8192$.}
     \label{Fig:SM_Phi_doping}
\end{figure*}

We also represent the doping dependence of pairing ITP $\beta \Phi_{yy}(\bold{q}, \beta/2)$ at fixed cylinder width $W = 6$ and temperature $T = 0.053$ in Fig.~\ref{Fig:SM_Phi_doping}. The dSC vs. PDW behavior at opposite doping sides is insensitive to the detailed doping levels as long as it is not too high. At overdoped regime, we find the $(\pi, \pi)$-pairing becomes dominant at hole-doped side and enhanced at electron-doped side, but the overall pairing fluctuations are suppressed upon further doping, which is consistent with the existence of an optimal doping level as anticipated.

\end{document}